\documentclass{article}

\usepackage{amssymb}
\usepackage{amsmath}

\begin{document}

\title{Quantum tomography for Dirac spinors}
\author{R. A. Mosna$^{(1,2)}$\thanks{E-mail address: mosna@ifi.unicamp.br}
\ and J. Vaz Jr$^{(2)}$\thanks{E-mail address: vaz@ime.unicamp.br}\\
{\small (1) Instituto de F\'{\i}sica Gleb Wataghin, Universidade Estadual de Campinas,}\\
{\small CP 6165, 13083-970, Campinas, SP, Brazil.}\\
{\small (2) Departamento de Matem\'{a}tica Aplicada, Universidade Estadual de Campinas,}\\
{\small CP 6065, 13081-970, Campinas, SP, Brazil.}}
\maketitle

\begin{abstract}
We present a tomographic scheme, based on spacetime symmetries, for the
reconstruction of the internal degrees of freedom of a Dirac spinor. We
discuss the circumstances under which the tomographic group can be taken as
$SU(2)$, and how this crucially depends on the choice of the gamma matrix
representation. A tomographic reconstruction process based on discrete
rotations is considered, as well as a continuous alternative.



\end{abstract}

\section{Introduction}
\label{sect:intro}

There is a long quest on the search of classical-like descriptions of quantum
mechanics. As examples, we can mention the approaches of Wigner \cite{wigner},
Moyal \cite{moyal}, Feynman \cite{feynman} and the various tentative hidden
variable theories. In the first two cases, a set of (possibly negative)
quasiprobability distributions, defined on the phase space, are the basic
variables of the theory. In Feynman's approach, negative probabilities are
allowed as a way to avoid the use of (probability) amplitudes. On the other
hand, the tomographic formulation of quantum mechanics \cite{bertrand,vogel,
mancini96,symplectic1,symplectic2,dodonov,manko,leonhardt,cassinelli,
mancini01,terra,weigert,relativistic} has received
considerable attention in recent years. In such an approach, the dynamical
variables of the theory are a set of probability distributions, which have
truly classical-like characteristics: they are non-negative, normalized and,
in principle, all measurable.

For a review on the principles of the tomographic approach, we refer the
reader to \cite{mancini96,mancini01}. Here we briefly outline its main
ideas. Consider a state $|\psi\rangle$ in some Hilbert space $\mathcal{H}$
describing a physical system. Let $\{|v_{\alpha}\rangle\}$ be an orthonormal 
basis of $\mathcal{H}$, whose elements are eigenvectors of a commuting set of 
Hermitian operators. Here, $\alpha$ should be interpreted as a multi-index 
that might contain discrete and/or continuous indices. Expanding 
$|\psi\rangle$ in this basis, we have 
$|\psi\rangle=\sum_{\alpha}\psi_{\alpha}|v_{\alpha}\rangle$, where the 
complex coefficients $\psi_{\alpha}$ represent probability amplitudes. 
The corresponding probabilities
\[
w_{\alpha}=|\psi_{\alpha}|^{2}=|\langle v_{\alpha}|\psi\rangle|^{2}
\]
are called \emph{marginal} distributions. Note that $w_{\alpha}$ are
non-negative normalized probabilities which are, in principle, all measurable.
The essence of the tomographic approach is to describe the physical state and
its dynamics in terms of the marginals.

Of course, the information relative to the phases in $\psi_{\alpha}$ is lost
when we consider the above marginals. Nevertheless, one can consider the
action on $\mathcal{H}$ of a family of transformations $U(g)$, labeled by a
certain parameter $g$ belonging to a (Lie) group $G$. Defining the
``rotated'' marginals $w_{\alpha}(g)=|\langle v_{\alpha}|U(g)|\psi\rangle|^{2}$ 
(which are again measurable in principle) and writing 
$\psi_{\alpha}(g)=\sum_{\beta}U(g)_{\alpha}^{\beta}\psi_{\beta}$,
it follows that the expression of $w_{\alpha}(g)=|\psi_{\alpha}(g)|^{2}$
carries interference terms among the relative phases of $\psi_{\beta}$. As a
result, one can find such relative phases in terms of the rotated marginals.

The tomographic schemes are usually written in terms of the density matrix $\rho$ 
associated with the physical system. Although in this work we are mainly interested
in pure states, this leads to a natural framework to study more general mixed
states. Then, the reconstruction process can be implemented by an integral
transformation $\rho=\int d\alpha dg$\thinspace$w_{\alpha}(g)K(\alpha,g)$, which 
determines $\rho$ in terms of the rotated marginals (when $\alpha$ is a discrete index, 
the corresponding integral should be replaced by a discrete sum). Some applications
of this tomographic scheme can be found in \cite{vogel}\ (optical tomography,
with $G=O(2)$), \cite{symplectic1,symplectic2,mancini01}\ (symplectic
tomography, with $G=Sp(2,\mathbb{R})$) and
\cite{dodonov,manko,mancini01,terra} (spin tomography, with
$G=SU(2)$). In \cite{weigert}, the interesting problem of defining a minimum
quorum of expectation values for the state reconstruction was addressed. Also,
a study of the properties of marginal distributions under relativistic
transformations, especially in the context of the relativistic oscillator
model, was presented in \cite{relativistic}.

In this work, we present a tomographic scheme for the reconstruction of the
internal degrees of freedom of Dirac spinors. These objects are known to
describe relativistic spin-1/2 particles, as electrons. More precisely, a
Dirac spinor 
$|\psi\rangle=(\psi_{1} \ \psi_{2} \ \psi_{3} \ \psi_{4})^{t}\in\mathbb{C}^{4}$ 
is an object carrying the representation
$D^{(1/2,0)}\oplus D^{(0,1/2)}$ of $Spin_{1,3}^{e}\cong Sl(2,\mathbb{C})$, the
covering group of the restricted Lorentz group. As $|\psi\rangle\in
\mathbb{C}^{4}$, a tomographic scheme based on $SU(4)$ would certainly work
for this case. However, to parallel the discussion with the non-relativistic
case, and to give a direct physical meaning to the transformations $U(g)$, we
demand that the tomographic group $G$ be \emph{generated by spacetime
transformations}.

The choice of the gamma matrix representation in the Dirac theory plays a
decisive role in this context. In fact, consider the tomographic
reconstruction of a generic Dirac spinor $|\psi\rangle$, with 7 degrees of
freedom (discounting a global phase). Let $\mathcal{L}$ be the restricted
Lorentz group and $\tilde{\mathcal{L}}$ the associated covering group. Then,
as we will show later,

(i) in the context of the Majorana representation, $|\psi\rangle$ can be
tomographically recovered by taking $G$ as the $SU(2)$ rotation subgroup 
of $\tilde{\mathcal{L}}$;

(ii) in the context of the standard representation, $|\psi\rangle$ can be
tomographically recovered if we take $G=\tilde{\mathcal{L}}$, but not for 
$G=SU(2)$ as in (i);

(iii) in the context of the chiral representation, $|\psi\rangle$ cannot be
tomographically reconstructed via spacetime transformations, i.e., even if 
$G$ is taken as the whole $\tilde{\mathcal{L}}$ (unless $|\psi\rangle$ is a 
Weyl spinor, corresponding to a massless particle).

It is a well known result that the Lorentz group is a non-compact space which
does not admit finite-dimensional unitary representations (except for the
trivial one) \cite{wigner:irep}. This means that boosts inevitably give rise
to non-unitary transformations for the spinor space. Although one might live
with this situation, it is clearly preferable to work only with rotations, if
possible. We see from the discussion above that, among the most common choices
for $\{\gamma^{\mu}\}$, namely the Majorana, standard and chiral
representations, only the first one is compatible with a tomographic procedure
based on spatial transformations. In this case, the tomographic group is given
by $SU(2)$. Alternatively, it is also possible to combine the marginals
associated with both the standard and chiral representations, so that a
tomographic reconstruction based on rotations is similarly achieved.

It should be noted that the discussion above regards the tomographic
reconstruction of a full Dirac spinor. If one wants to reconstruct a spinor
that is already known to be in the positive energy sector, then clearly less
symmetry transformations are required (however, it is well known that this
sector is not preserved by time evolution \cite{ItzyksonZuber80}; this is, in
fact, a problem of the first quantized Dirac theory).

We observe that a shortcoming of describing a Dirac spinor by means of the
marginals $w_{k}=|\psi_{k}|^{2}$, when compared to the non-relativistic case,
is the lack of a clear interpretation for these objects. In fact, consider the
case of the standard representation. The projector associated with positive
energy and spin up (in the $z$-direction) then reduces to
\[
\frac{1}{2}(I+\gamma_{0})\frac{1}{2}(I+i\gamma_{1}\gamma_{2})=
\left(
\begin{array}{cccc}
1 & 0 & 0 & 0\\
0 & 0 & 0 & 0\\
0 & 0 & 0 & 0\\
0 & 0 & 0 & 0
\end{array}
\right)  ,
\]
only in the reference frame in which the particle is at rest. When we consider
an arbitrary particle in an indefinite state, we must consider the Fourier
expansion of $\psi(x)$ in terms of eigenstates of momentum $\phi(p)$. But
then, the physical interpretation of the first component of $\phi(p)$ changes
with $p$. This happens because boosts mix the components $\psi_{k}$ of
$|\psi\rangle$. Therefore, the marginals $w_{k}=|\psi_{k}|^{2}$ do
not correspond to an easy-to-describe physical property of the 
particle.\footnote{We note that this
situation does not occur in Pauli theory, for a non-relativistic
boost (i.e., a Galileo transformation corresponding to a change of velocity
between frames) does not mix the components of a Pauli spinor.} When we
consider alternative gamma matrix representations, the interpretation of the
marginals are even more unclear.

On the other hand, the bilinear covariants associated with $|\psi\rangle$ provide
another classical-like description of the Dirac theory, in the sense that they are,
in principle, measurable tensorial densities.\footnote{The bilinear covariant
$J^{\mu}=\langle\bar{\psi}| \gamma^{\mu}| \psi\rangle$ corresponds to the charge density
($eJ^{0}$) and electric current density ($ecJ^{k}$) associated with a Gibbs ensemble
of identical particles; $S^{\mu\nu}= \langle\bar{\psi}|i\gamma^{\mu\nu}|\psi\rangle$ 
corresponds to magnetic ($\tfrac{e\hbar}{2mc} S^{ij}$) and electric 
($\tfrac{e\hbar}{2mc} S^{0j}$) moment densities; $\tfrac{\hbar}{2} K^{\mu} = 
\tfrac{\hbar}{2} \langle\bar{\psi}|i\gamma_{0123}\gamma^{\mu}|\psi\rangle$
corresponds to the spin density. The scalar and pseudo-scalar bilinear covariants have
a less clear interpretation, but the Fierz identities (see section \ref{sect:bilcov})
can be used to express them in terms of $J^{\mu}$, $S^{\mu\nu}$ and $K^{\mu}$ 
\cite{deBroglie,takabayasi} (see also \cite{baym,hollandbook}).} Moreover, it is
natural to expect that the manifest covariance of these quantities should somehow 
favor them over the marginals. For this reason, our plan in this article is to 
establish a well defined correspondence between the bilinear covariants and the 
marginals $w_{k}$, so that the latter inherit the measurability (and possibly the 
dynamics) of the former. As a result, in spite of the above difficulties, the 
quantities $w_{k}$ do provide a set of non-negative, normalized and measurable 
quantities, describing the quantum state of the particle. Furthermore, it is known 
that the Dirac theory can be formulated in terms of the bilinear covariants 
\cite{crawfordproc} (see also \cite{holland}). Then, the above procedure might be 
useful for obtaining a Dirac equation written in terms of tomographic and 
classical-like quantities like $w_{k}$.\footnote{Note, however, that this would 
still require the reconstruction of the (position-dependent) global phase of 
$|\psi\rangle$.}

This article is organized as follows. In section \ref{sect:bilcov}, we review
some facts about Dirac spinors, including a reconstruction theorem
\cite{crawford85} that allows one to obtain $|\psi\rangle$ from the bilinear
covariants. In section \ref{sect:marginals}, we obtain the aforementioned
correspondence between the bilinear covariants and the marginals $w_{k}$. 
This can be considered a generalization of (the non-relativistic) spin 
tomography techniques presented in \cite{dodonov,manko,mancini01,terra}.
The covariance of the bilinear covariants is then explored to tomographically 
reconstruct the spinor $|\psi\rangle$ from the marginals. In 
section \ref{sect:gamma}, we discuss the dependence of this 
tomographic approach on the choice of the gamma matrix representation. 
Section \ref{sect:conclusion} is reserved for some final remarks. In what 
follows, we use natural units ($\hbar=c=1$).

\section{Bilinear covariants}
\label{sect:bilcov}

In order to establish notation, let us briefly review some well known facts
about the Dirac theory. Let 
$|\psi\rangle=(\psi_{1} \ \psi_{2} \ \psi_{3} \ \psi_{4})^{t}$ be a Dirac 
spinor (in what follows, $|\psi\rangle$ always represents a pure state). 
Under a restricted Lorentz transformation $\Lambda=(\Lambda^{\mu}_{\phantom{\mu}\nu})$, 
this object transforms as $|\psi^{\prime}(x)\rangle=L|\psi(\Lambda^{-1}x)\rangle$, 
where the matrix $L$ is related to $\Lambda^{\mu}_{\phantom{\mu}\nu}$ by
$L^{-1}\gamma^{\mu}L=\Lambda^{\mu}_{\phantom{\mu}\nu}\gamma^{\nu}$ 
\cite{ItzyksonZuber80} ($L$ belongs to the covering space $\tilde{\mathcal{L}}$ 
of the restricted Lorentz group \cite{lounesto}). Denoting the Dirac 
conjugate of $|\psi\rangle$ by $\langle\bar{\psi}|=\langle\psi|\gamma_{0}$, 
we then have $\langle\bar{\psi}^{\prime}|=\langle\bar{\psi}|L^{-1}$. This 
immediately yields the following 16 tensorial quantities, known as bilinear 
covariants:
\begin{align}
\Omega_{1} &  =\langle\bar{\psi}|\psi\rangle,\nonumber\\
J^{\mu} &  =\langle\bar{\psi}|\gamma^{\mu}|\psi\rangle,\nonumber\\
S^{\mu\nu} &  =\langle\bar{\psi}|i\gamma^{\mu\nu}|\psi\rangle,
\textrm{ with }\gamma^{\mu\nu}=\tfrac{1}{2}[\gamma^{\mu},\gamma^{\nu}],\label{bilcov}\\
K^{\mu} &  =\langle\bar{\psi}|i\gamma_{0123}\gamma^{\mu}|\psi\rangle,
\text{ with }\gamma_{0123}=\gamma_{0}\gamma_{1}\gamma_{2}\gamma_{3},\nonumber\\
\Omega_{2} &  =\langle\bar{\psi}|-\gamma_{0123}|\psi\rangle,\nonumber
\end{align}
which transform respectively as a scalar, a 4-vector, a tensor of second
degree, a pseudo-vector and a pseudo-scalar. These quantities are obviously not
independent. The constraint relations among them are given by the Fierz
identities (we use the conventions of \cite{lounesto}):
\begin{subequations}
\label{Fierzident}
\begin{align}
J_{\mu}J^{\mu} &  =\Omega_{1}^{2}+\Omega_{2}^{2},\label{Fierzid.a}\\
J_{\mu}J^{\mu} &  =-K_{\mu}K^{\mu},\label{Fierzid.b}\\
J_{\mu}K^{\mu} &  =0,\label{Fierzid.c}\\
J_{\mu}K_{\nu}-K_{\mu}J_{\nu} &  =-(\Omega_{2}S_{\mu\nu}+\Omega_{1}
\tfrac{1}{2}\epsilon_{\mu\nu\alpha\beta}S^{\alpha\beta}),\label{Fierzid.d}
\end{align}
\end{subequations}
with $\epsilon_{0123}=1$. These 9 equations (note the anti-symmetry in 
$\mu\nu$) reduce the number of independent bilinear covariants to 7, as expected.
Indeed, $|\psi\rangle$ has eight real components which reduce to seven
independent quantities when a global phase is discarded.

Let us denote a bilinear covariant generically by 
$\rho^{a}=\langle\bar{\psi}|\Gamma^{a}|\psi\rangle$, where $\Gamma^{a}$ 
can be read from eqs. (\ref{bilcov}) (we note that the 16 matrices $\Gamma^{a}$ 
form a basis for the Dirac algebra). As the bilinear covariants are real, the 
quantity
\begin{equation}
\rho:={\textstyle\sum_{a}} \, \rho^{a}\Gamma_{a}
\label{def:rho}
\end{equation}
can be thought of as a vector in a 16-dimensional real vector space. In this
context, the Fierz identities determine a 7-dimensional submanifold of
$\mathbb{R}^{16}$ in which $\rho$ lives in \cite{crawford90}. This
submanifold generalizes the Bloch sphere of Pauli theory.

It follows from eq. (\ref{def:rho}) that
\[
\rho=\Omega_{1}+J+iS+iK\gamma_{0123}+\Omega_{2}\gamma_{0123},
\]
where $J=J^{\mu}\gamma_{\mu}$, $S=\tfrac{1}{2}S^{\mu\nu}\gamma_{\mu\nu}$ and
$K=K^{\mu}\gamma_{\mu}$. It is also useful to note that there is a natural
inner product defined on the Dirac algebra by 
$(A,B)=(1/4)\operatorname{tr}(A\bar{B}) $, where 
$\bar{B}:=\gamma_{0}B^{\dag}\gamma_{0}$ is the Dirac conjugate of $B $. Note that 
$\rho^{a}=\langle\bar{\psi}|\Gamma^{a}|\psi\rangle=
\operatorname{tr}\left(  \Gamma^{a}|\psi\rangle\langle\bar{\psi}|\right) =
4\left(  \Gamma^{a},|\psi\rangle\langle\bar{\psi}|\right)  ,$ so that we can 
alternatively write
\begin{equation}
\rho=4|\psi\rangle\langle\bar{\psi}|. \label{rho}
\end{equation}

A reconstruction theorem \cite{crawford90} can be used to obtain
$|\psi\rangle$, apart from a global phase, from the bilinear covariants, i.e.,
from $\rho$. To see this, consider the action of $\rho$ on a fixed spinor
$|\eta\rangle$ (usually taken as $|\eta\rangle=(1 \ 0 \ 0 \ 0)^{t}$). This 
gives $\rho|\eta\rangle=4|\psi\rangle\langle\bar{\psi}|\eta\rangle$, 
and so $|\psi\rangle=\frac{1}{4\langle\bar{\psi}|\eta\rangle}\rho|\eta\rangle$. As 
$\langle\bar{\eta}|\rho|\eta\rangle=4|\langle\bar{\psi}|\eta\rangle|^{2}$, we have 
$\langle \bar{\psi}|\eta\rangle=e^{i\phi}|\langle\bar{\psi}|\eta\rangle|=
e^{i\phi}\left(  \tfrac{1}{4}\langle\bar{\eta}|\rho|\eta\rangle\right)  ^{1/2}$.
Substitution in the expression for $|\psi\rangle$ yields
\[
|\psi\rangle=\frac{e^{-i\phi}}{\sqrt{4\langle\bar{\eta}|\rho|\eta\rangle}}\rho|\eta\rangle.
\]
This can be brought to a simpler form if we rescale the bilinear covariants by
$R=\omega\rho$, with $\omega=(4\langle\bar{\eta}|\rho|\eta\rangle)^{-1/2}$. As
a result, we can write
\[
|\psi\rangle=e^{-i\phi}R|\eta\rangle.
\]

\section{Marginal distributions and bilinear covariants}
\label{sect:marginals}

Let us now relate the marginal distributions to the bilinear covariants. To 
do that, we expand $|\psi\rangle$ in terms of the canonical basis 
$\{|v_{k}\rangle\}_{k=1}^{4}$ of $\mathbb{C}^{4}$, with 
$|v_{1}\rangle=(1 \ 0 \ 0 \ 0)^{t}$, $|v_{2}\rangle=(0 \ 1 \ 0 \ 0)^{t}$ and 
so on. If $|\psi\rangle=\sum\psi_{k}|v_{k}\rangle$, we have
\[
w_{k}=|\psi_{k}|^{2},\quad k=1,2,3,4.
\]
Consider the matrices (or projection operators) given by
\begin{equation}
P_{1}=\left(
\begin{array}{cccc}
1 & 0 & 0 & 0\\
0 & 0 & 0 & 0\\
0 & 0 & 0 & 0\\
0 & 0 & 0 & 0
\end{array}
\right)  , \ 
P_{2}=\left(
\begin{array}{cccc}
0 & 0 & 0 & 0\\
0 & 1 & 0 & 0\\
0 & 0 & 0 & 0\\
0 & 0 & 0 & 0
\end{array}
\right)  , \
P_{3}=\left(
\begin{array}{cccc}
0 & 0 & 0 & 0\\
0 & 0 & 0 & 0\\
0 & 0 & 1 & 0\\
0 & 0 & 0 & 0
\end{array}
\right)  , \
P_{4}=\left(
\begin{array}{cccc}
0 & 0 & 0 & 0\\
0 & 0 & 0 & 0\\
0 & 0 & 0 & 0\\
0 & 0 & 0 & 1
\end{array}
\right)  .\label{Pk}
\end{equation}
Then
\begin{equation}
w_{k}=\langle\psi|P_{k}|\psi\rangle=\langle\bar{\psi}|\gamma_{0}P_{k}|\psi\rangle,
\quad k=1,2,3,4.\label{expr.for.wk}
\end{equation}
To go further, we need to choose a specific representation of the gamma
matrices to work with. In the Majorana representation \cite{ItzyksonZuber80}
(other choices will be discussed shortly)
\[
\gamma_{0}^{mj}=\left(
\begin{array}{cc}
0 & \sigma_{2}\\
\sigma_{2} & 0
\end{array}
\right)  ,\quad
\gamma_{1}^{mj}=\left(
\begin{array}{cc}
-i\sigma_{3} & 0\\
0 & -i\sigma_{3}
\end{array}
\right)  ,\quad
\gamma_{2}^{mj}=\left(
\begin{array}{cc}
0 & \sigma_{2}\\
-\sigma_{2} & 0
\end{array}
\right)  ,\quad
\gamma_{3}^{mj}=\left(
\begin{array}{cc}
i\sigma_{1} & 0\\
0 & i\sigma_{1}
\end{array}
\right)  ,
\]
where $\sigma_{k}$ are the Pauli matrices, a straightforward calculation yields
\begin{equation}
\begin{tabular}{l}
$P_{1}=\frac{1}{2}(1+\gamma_{20}^{mj})\frac{1}{2}(1+i\gamma_{1}^{mj}),\qquad
P_{2}=\frac{1}{2}(1+\gamma_{20}^{mj})\frac{1}{2}(1-i\gamma_{1}^{mj}),$\\
$P_{3}=\frac{1}{2}(1-\gamma_{20}^{mj})\frac{1}{2}(1+i\gamma_{1}^{mj}),\qquad
P_{4}=\frac{1}{2}(1-\gamma_{20}^{mj})\frac{1}{2}(1-i\gamma_{1}^{mj}).$
\end{tabular}
\label{Pk.for.Majorana}
\end{equation}
Substitution in eq. (\ref{expr.for.wk}) leads to 
$w_{mj,k}=\langle\bar{\psi}|\tfrac{1}{4}(\gamma_{0}^{mj}\pm\gamma_{2}^{mj}\pm 
i\gamma_{01}^{mj}\pm i\gamma_{12}^{mj})|\psi\rangle=
\tfrac{1}{4}(J_{0}\pm J_{2}\pm S_{01}\pm S_{12})$, where the $\pm$ signs 
vary with $k$. More precisely
\[
\begin{tabular}{l}
$w_{mj,1}=\tfrac{1}{4}\left(  J_{0}-J_{2}+S_{01}+S_{12}\right)  ,$\\
$w_{mj,2}=\tfrac{1}{4}\left(  J_{0}-J_{2}-S_{01}-S_{12}\right)  ,$\\
$w_{mj,3}=\tfrac{1}{4}\left(  J_{0}+J_{2}+S_{01}-S_{12}\right)  ,$\\
$w_{mj,4}=\tfrac{1}{4}\left(  J_{0}+J_{2}-S_{01}+S_{12}\right)  .$
\end{tabular}
\]
After solving for $J_{0},J_{2},S_{01}$ and $S_{12}$, we have
\begin{equation}
\begin{tabular}{l}
$J_{0}=w_{mj,1}+w_{mj,2}+w_{mj,3}+w_{mj,4},$\\
$J_{2}=-w_{mj,1}-w_{mj,2}+w_{mj,3}+w_{mj,4},$\\
$S_{01}=w_{mj,1}-w_{mj,2}+w_{mj,3}-w_{mj,4},$\\
$S_{12}=w_{mj,1}-w_{mj,2}-w_{mj,3}+w_{mj,4}.$
\end{tabular}
\label{bilcov.from.marginals}
\end{equation}
The result is a partial recovering of $\rho$ in terms of the marginal
distributions. Now we explore the symmetries of the Lorentz group to obtain
the full expression for $\rho$, still in terms of marginal distributions.

Applying a restricted Lorentz transformation 
$\Lambda=(\Lambda^{\mu}_{\phantom{\mu}\nu})$
to our system, we have $|\psi^{\prime}\rangle=L|\psi\rangle$. The corresponding 
$\Lambda$-dependent marginal distributions (cf eq. (\ref{expr.for.wk})) are:
\begin{equation}
w_{k}^{(\Lambda)}=\langle\bar{\psi}^{\prime}|\gamma_{0}P_{k}|\psi^{\prime}\rangle=
\langle\bar{\psi}|L^{-1}\gamma_{0}P_{k}L|\psi\rangle.\label{wkLambda}
\end{equation}
On the other hand, the new bilinear covariants are given by 
$\Omega_{1}^{(\Lambda)}=\Omega_{1}$, 
$J_{\mu}^{(\Lambda)}=\Lambda_{\mu}^{\phantom{\mu}\nu}J_{\nu}$,
$S_{\mu\nu}^{(\Lambda)}=\Lambda_{\mu}^{\phantom{\mu}\alpha}
\Lambda_{\nu}^{\phantom{\nu}\beta}S_{\alpha\beta}$,
$K_{\mu}^{(\Lambda)}=\Lambda_{\mu}^{\phantom{\mu}\nu}K_{\nu}$ and
$\Omega_{2}^{(\Lambda)}=\Omega_{2}$
(in the above notation, $w_{mj,k}=w_{mj,k}^{(I)}$, $J_{\mu}=J_{\mu}^{(I)}$, and so on
denote the quantities associated with the original frame, before the application of the 
symmetry transformation). It follows that 
$w_{mj,k}^{(\Lambda)}=\langle\bar{\psi}|L^{-1}\tfrac{1}{4}
(\gamma_{0}^{mj}\pm\gamma_{2}^{mj}\pm i\gamma_{01}^{mj}\pm i\gamma_{12}^{mj})
L|\psi\rangle=\tfrac{1}{4}
(J_{0}^{(\Lambda)}\pm J_{2}^{(\Lambda)}\pm S_{01}^{(\Lambda)}\pm S_{12}^{(\Lambda)})$, 
where the $\pm$ signs vary with $k$. More precisely,
\begin{equation}
\begin{tabular}{l}
$J_{0}^{(\Lambda)}=w_{mj,1}^{(\Lambda)}+w_{mj,2}^{(\Lambda)}+w_{mj,3}^{(\Lambda)}+
w_{mj,4}^{(\Lambda)},$\\
$J_{2}^{(\Lambda)}=-w_{mj,1}^{(\Lambda)}-w_{mj,2}^{(\Lambda)}+w_{mj,3}^{(\Lambda)}+
w_{mj,4}^{(\Lambda)},$\\
$S_{01}^{(\Lambda)}=w_{mj,1}^{(\Lambda)}-w_{mj,2}^{(\Lambda)}+w_{mj,3}^{(\Lambda)}-
w_{mj,4}^{(\Lambda)},$\\
$S_{12}^{(\Lambda)}=w_{mj,1}^{(\Lambda)}-w_{mj,2}^{(\Lambda)}-w_{mj,3}^{(\Lambda)}+
w_{mj,4}^{(\Lambda)}.$
\end{tabular}
\label{bilcovLambda.Maj.from.marginals}
\end{equation}
Now we can vary $\Lambda$ in the above expressions to recover all the bilinear covariants:

(a) taking $\Lambda=I=$[identity] (i.e. no symmetry transformation), we
determine (from eqs. (\ref{bilcovLambda.Maj.from.marginals}) or
eqs. (\ref{bilcov.from.marginals})) $J_{0},J_{2},S_{01}$ and $S_{12}$ 
in terms of $w_{mj,k}$;

(b) taking $\Lambda=R_{x}=$[$\pi/2$-rotation about the $x$-axis], we have
$J_{0}=J_{0}^{(R_{x})},S_{01}=S_{01}^{(R_{x})}$ and $J_{3}=J_{2}^{(R_{x})},
S_{31}=-S_{12}^{(R_{x})}$. All these quantities are determined by the marginals 
$w_{mj,k}^{(R_{x})}$ from eqs. (\ref{bilcovLambda.Maj.from.marginals});

(c) taking $\Lambda=R_{y}=$[$\pi/2$-rotation about the $y$-axis], we analogously
obtain $J_{0}=J_{0}^{(R_{y})},J_{2}=J_{2}^{(R_{y})}$ and 
$S_{03}=-S_{01}^{(R_{y})},S_{23}=S_{12}^{(R_{y})}$.  All these quantities are 
determined by the marginals $w_{mj,k}^{(R_{y})}$ from 
eqs. (\ref{bilcovLambda.Maj.from.marginals});

(d) taking $\Lambda=R_{z}=$[$\pi/2$-rotation about the $z$-axis], we analogously
obtain $J_{0}=J_{0}^{(R_{z})},S_{12}=S_{12}^{(R_{z})}$ and 
$J_{1}=-J_{2}^{(R_{z})},S_{02}=S_{01}^{(R_{z})}$.  All these quantities are 
determined by the marginals $w_{mj,k}^{(R_{z})}$ from 
eqs. (\ref{bilcovLambda.Maj.from.marginals}).

So far, we have recovered the ten bilinear covariants $J_{\mu}$ and $S_{\mu\nu}$. 
The rest of them, namely $\Omega_{1}$, $\Omega_{2}$ and $K_{\mu}$, are
easily obtained from the Fierz identities. In fact, eqs. (\ref{Fierzident}) 
yield the identities \cite{crawford85}
\[
\Omega_{1}K_{\nu}=J^{\mu}(\ast S)_{\mu\nu},\qquad
\Omega_{2}K_{\nu}=-J^{\mu}S_{\mu\nu},
\]
where $(\ast S)_{\mu\nu}=-\tfrac{1}{2}\epsilon_{\mu\nu\alpha\beta}S^{\alpha\beta}$, 
with $\epsilon_{0123}=1$. In this way, all the bilinear covariants are obtained 
from the rotated marginals.

Writing everything in terms of the rotated marginals, it follows from (a)-(d)
above and eqs. (\ref{bilcovLambda.Maj.from.marginals}) that:
\begin{equation}
\begin{tabular}{l}
$J_{0}=w_{mj,1}+w_{mj,2}+w_{mj,3}+w_{mj,4},$\\
$J_{1}=w_{mj,1}^{(R_{z})}+w_{mj,2}^{(R_{z})}-w_{mj,3}^{(R_{z})}-w_{mj,4}^{(R_{z})},$\\
$J_{2}=-w_{mj,1}-w_{mj,2}+w_{mj,3}+w_{mj,4},$\\
$J_{3}=-w_{mj,1}^{(R_{x})}-w_{mj,2}^{(R_{x})}+w_{mj,3}^{(R_{x})}+w_{mj,4}^{(R_{x})},$\\
$S_{01}=w_{mj,1}-w_{mj,2}+w_{mj,3}-w_{mj,4},$\\
$S_{02}=w_{mj,1}^{(R_{z})}-w_{mj,2}^{(R_{z})}+w_{mj,3}^{(R_{z})}-w_{mj,4}^{(R_{z})},$\\
$S_{03}=-w_{mj,1}^{(R_{y})}+w_{mj,2}^{(R_{y})}-w_{mj,3}^{(R_{y})}+w_{mj,4}^{(R_{y})},$\\
$S_{12}=w_{mj,1}-w_{mj,2}-w_{mj,3}+w_{mj,4},$\\
$S_{23}=w_{mj,1}^{(R_{y})}-w_{mj,2}^{(R_{y})}-w_{mj,3}^{(R_{y})}+w_{mj,4}^{(R_{y})},$\\
$S_{31}=-w_{mj,1}^{(R_{x})}+w_{mj,2}^{(R_{x})}+w_{mj,3}^{(R_{x})}-w_{mj,4}^{(R_{x})}.$
\end{tabular}
\label{bilcov.from.rotated.marginals}
\end{equation}
As we mentioned above, these quantities determine all the bilinear covariants,
and thus reconstruct the spinor $| \psi \rangle $ as in the
previous section. Moreover, the 6 relations 
$J_{0}=J_{0}^{(R_{x})}=J_{0}^{(R_{y})}=J_{0}^{(R_{z})},
S_{01}=S_{01}^{(R_{x})},J_{2}=J_{2}^{(R_{y})},S_{12}=S_{12}^{(R_{z})}$ in (b)-(d) 
yield 6 constraint equations among the 16 marginals above. This can be used to 
reduce the number of marginals in eqs. (\ref{bilcov.from.rotated.marginals}).

It is important to note that, in the above reconstruction of $|\psi\rangle$,
we did not employ boosts. Indeed, the relevant tomographic group was
generated by the $\Lambda$'s in the rotation subgroup $SO(3)$ of the Lorentz
group $\mathcal{L}$. This corresponds to elements $L$ in a $SU(2)$ subgroup of
the associated covering group 
$\tilde{\mathcal{L}}=Spin_{1,3}^{e}\cong Sl(2,\mathbb{C})$.

\subsection{A continuous alternative}

From our previous discussion, we see that a crucial step to the tomographic
recovering process is to reconstruct a vector $\mathbf{v}\in\mathbb{R}^{3}$ if
one of its components is known in all frames. Let us fix a reference frame $K$
and let us denote the $K$-components of $\mathbf{v}$ by $(v^{1},v^{2},v^{3})$.
Suppose we know the third component of $\mathbf{v}$ in all frames. Given
another reference frame $K^{\prime}$, if $\theta$ and $\varphi$ are the polar
and azimuthal angles of $\mathbf{e}_{3}^{\prime}$ in relation to $K$, we have
\[
\mathbf{e}_{3}^{\prime}(\theta,\varphi)=
\sin\theta\cos\varphi \ \mathbf{e}_{1}+
\sin\theta\sin\varphi \ \mathbf{e}_{2}+
\cos\theta \ \mathbf{e}_{3}.
\]
Let $\nu(\theta,\varphi)$ be the third $K^{\prime}$-component of $\mathbf{v}$,
i.e. $\nu(\theta,\varphi)=\mathbf{v}\cdot\mathbf{e}_{3}^{\prime}$. Then, we
can reconstruct $\mathbf{v}$ from $\nu(\theta,\varphi)$ by at least two procedures:

\paragraph{(I) Discrete method}

As $\mathbf{e}_{3}^{\prime}(\pi/2,0)=\mathbf{e}_{1}$, 
$\mathbf{e}_{3}^{\prime}(\pi/2,\pi/2)=\mathbf{e}_{2}$ and 
$\mathbf{e}_{3}^{\prime}(0,0)=\mathbf{e}_{3}$ we have 
$v^{1}=\nu(\pi/2,0)$, $v^{2}=\nu(\pi/2,\pi/2)$ and $v^{3}=\nu(0,0)$. Thus, 
$\mathbf{v}=\nu(\pi/2,0)\mathbf{e}_{1}+\nu(\pi/2,\pi/2)\mathbf{e}_{2}+
\nu(0,0)\mathbf{e}_{3}$. This is the reconstruction method
we used in the previous section.

\paragraph{(II) Continuous method}

This method goes along the lines of \cite{dodonov,manko}, in which all the
directions $(\theta,\varphi)$ are considered. The idea is to recover
$\mathbf{v}$ by an integral transformation of $\nu(\theta,\varphi):$
$\mathbf{v}=\int_{S^{2}}d\Omega\mathbf{A}(\theta,\varphi)\nu(\theta,\varphi)$,
where $d\Omega=\sin\theta d\theta d\varphi$ is the solid angle element on 
the sphere. There is a lot of ambiguity in choosing the kernel 
$\mathbf{A}(\theta,\varphi)$, but a simple choice is given by 
$\mathbf{A}(\theta,\varphi)=
\left(  
\tfrac{2}{\pi^{2}}\cos\varphi,\tfrac{2}{\pi^{2}}\sin\varphi,\tfrac{3}{4\pi}\cos\theta
\right)$.\bigskip

Each of the methods above lead to a different set of
tomographic quantities describing the spinor. In the previous section, we
employed a discrete method. On the other hand, if the continuous method is
employed, the spinor would be described in terms of continuous variables
analogous to $\nu(\theta,\varphi)$.

\section{On the choice of the gamma matrix representation}
\label{sect:gamma}

In this section, we discuss the dependence of the tomographic approach
developed above on the choice of the gamma matrix representation. Consider the
expression (\ref{wkLambda}) for $w_{k}^{(\Lambda)}$ in terms of the
projection operators in eqs. (\ref{Pk}):
\begin{equation}
w_{k}^{(\Lambda)}=\langle\bar{\psi}|L^{-1}\gamma_{0}P_{k}L|\psi\rangle.
\label{wkLambda.2}
\end{equation}

Of course, the functional dependence of $P_{k}$ in terms of $\gamma_{\mu}$, 
$\mu=0,1,2,3$, depends on the particular choice of the gamma matrix
representation. For the Majorana representation, this is given by eqs.
(\ref{Pk.for.Majorana}). A straightforward calculation shows that the
analogous expressions for the standard and chiral representations are
\begin{subequations}
\label{Pk.for.st.and.ch}
\begin{align}
P_{k}^{st}  &  =\tfrac{1}{2}(1\pm\gamma_{0}^{st})
\tfrac{1}{2}(1\pm i\gamma_{12}^{st}),\label{Pk.st}\\
P_{k}^{ch}  &  =\tfrac{1}{2}(1\pm\gamma_{30}^{ch})
\tfrac{1}{2}(1\pm i\gamma_{0123}^{ch}),
\label{Pk.ch}
\end{align}
\end{subequations}
where the $\pm$ signs vary with $k$. It follows from eq. (\ref{wkLambda.2})
that
\begin{subequations}
\label{wkLambda.st.and.ch}
\begin{align}
w_{st,k}^{(\Lambda)}  &  =\tfrac{1}{4}\left[  \Omega_{1}^{(\Lambda)}\pm
J_{0}^{(\Lambda)}\pm S_{12}^{(\Lambda)}\pm K_{3}^{(\Lambda)}\right],
\label{wkLambda.st}\\
w_{ch,k}^{(\Lambda)}  &  =\tfrac{1}{4}\left[  J_{0}^{(\Lambda)}\pm
J_{3}^{(\Lambda)}\pm K_{0}^{(\Lambda)}\pm K_{3}^{(\Lambda)}\right]  .
\label{wkLambda.ch}
\end{align}
\end{subequations}

For the standard representation, we see from eq. (\ref{wkLambda.st})
that, by performing rotations, we can recover $\Omega_{1},J_{0},S_{23},
S_{31},S_{12},K_{1},K_{2}$ and $K_{3}$ from the marginals. Unfortunately,
these bilinear covariants apparently do not suffice to entirely recover a
generic Dirac spinor, with 7 degrees of freedom (discounting a global phase).
This is more easily seen from the form of the generators of rotations
(associated with the standard representation):
\[
i\gamma_{jk}^{st}=\frac{i}{2}[\gamma_{j}^{st},\gamma_{k}^{st}]=
\left(
\begin{array}{cc}
\sigma_{l} & 0\\
0 & \sigma_{l}
\end{array}
\right)  ,\quad (jkl)\textrm{ cyclic, }l=1,2,3.
\]
It follows that rotations do not mix the first two components (i.e. $\psi_{1}$ 
and $\psi_{2}$) of $|\psi\rangle=(\psi_{1} \ \psi_{2} \ \psi_{3} \ \psi_{4})^{t}$ 
with the last ones (i.e. $\psi_{3}$ and $\psi_{4}$). In this way,
the relative phase between the first and the last set of components of
$|\psi\rangle$ cannot be recovered solely with rotations. On the other hand,
if boosts were allowed to reconstruct the spinor, then we could also obtain
$J_{1},J_{2},J_{3}$ (from $J_{0}$), $S_{01},S_{02},S_{03}$ (from
$S_{23},S_{31},S_{12}$) and $K_{0}$ (from $K_{3}$). This would certainly
reconstruct the spinor. Therefore, we have shown that, in the standard
representation, the state can be reconstructed by means of restricted Lorentz
transformations, but not through rotations (cf introduction).

For the chiral representation, we see from eq. (\ref{wkLambda.ch}) that
even if the whole Lorentz group is used, we can only recover $J_{\mu}$ and
$K_{\mu}$, $\mu=0,1,2,3$, from the marginals. This is not sufficient to
reconstruct all the bilinear covariants for a generic Dirac spinor, with 7
degrees of freedom (discounting a global phase). In fact, we know from the
Fierz identities that $J^{\mu}J_{\mu}=-K^{\mu}K_{\mu}$ and $J^{\mu}K_{\mu}=0$,
and thus there are only 6 independent quantities in $J_{\mu}$ and $K_{\mu}$.
This situation changes if $|\psi\rangle$ is a Weyl spinor, corresponding to a
massless spin-1/2 particle. In that case, we always have 
$\Omega_{1}=\Omega_{2}=0$ and $S=0$ \cite{lounesto} and then the
reconstruction process could proceed as before.\footnote{We also note that 
\cite{lounesto} (i) for a Dirac spinor, $\Omega_{1}$ and $\Omega_{2}$ are not 
both zero, (ii) for a Weyl spinor, $\Omega_{1}=\Omega_{2}=0$ and $S_{\mu\nu}=0$ 
and (iii) for a Majorana spinor, $\Omega_{1}=\Omega_{2}=0$ and $K_{\mu}=0$.} 
Note that the marginals $w_{ch,k}$ are associated with the probability of finding 
the particle with positive/negative chirality and spin up/down (in the
$z$-direction). As a massless particle has definite chirality, the knowledge
of the $w_{ch,k}$, $k=1,2,3,4$, would be enough information to recover its
state. But, as we have seen, this is not true for a massive particle. This
shows our claim (see introduction) that, in the chiral representation, the
state of a massive particle cannot be reconstructed from spacetime symmetries, 
i.e., with $G$ contained in $\tilde{\mathcal{L}}$.

Therefore, unlike the Majorana representation, neither the standard nor the
chiral representations can be isolatedly used in such a tomographic scheme,
based on rotations, for a generic Dirac spinor. A way out of this
difficulty is to combine the marginals coming from both the standard and the
chiral transformations. We see from the above discussion that, by performing
rotations, we can then recover $\Omega_{1},J_{\mu},S_{23},S_{31},S_{12}$ and
$K_{\mu}$. These 12 bilinear covariants determine the rest of them through the
Fierz identities, and we can proceed as before.

\subsection*{The general case}

It is well known that an arbitrary representation $\{\gamma_{\mu}\}$ of the
gamma matrices can be written as 
$\gamma_{\mu}=U\gamma_{\mu}^{st}U^{-1}$, where 
$\gamma_{\mu}^{st}$ corresponds to the standard representation and 
$U$ is an unitary matrix. It follows from eq. 
(\ref{Pk.st}) that, in terms of the new gamma matrices 
$\{\gamma_{\mu}\}$:
\[
P_{k}=\frac{1}{2}(1\pm u)\frac{1}{2}(1\pm i\sigma),
\]
where $u=U^{-1}\gamma_{0}^{st}U$ and 
$\sigma=U^{-1}\gamma_{12}^{st}U$. The discussion above 
shows that a tomographic scheme, based on spacetime transformations, works 
for the representation $\{\gamma_{\mu}\}$ only if the bilinear covariants 
associated with $\gamma_{0}$, $\gamma_{0}u$, $\gamma_{0}\sigma$ and 
$\gamma_{0}u\sigma$ are independent enough to reconstruct the spinor.
It may happen that such a reconstruction is possible using only spatial 
rotations (as in the Majorana representation), or using necessarily boosts 
and rotations (as in the standard representation), or even not possible inside 
the Lorentz group (as in the chiral representation, with massive particles).

\section{Concluding remarks}
\label{sect:conclusion}

We have presented a tomographic scheme, based on spacetime transformations,
for the reconstruction of the internal degrees of freedom of a Dirac spinor.
The assumption that the tomographic group $G$ is generated by spacetime
transformations was shown to restrict the choice of the gamma matrices. The
cases of standard, chiral and Majorana representations were studied in detail.
We also analyzed under what conditions $G$ can be taken as $SU(2)$. A direct
tomographic process based on discrete rotations was considered, as well as a 
continuous alternative. Finally, as we mentioned in the introduction, the
method considered here might be useful for obtaining an analogue of the Dirac
equation in terms of tomographic quantities.\bigskip

\noindent\textbf{Acknowledgments }\textit{The authors are grateful to M. O.
Terra Cunha and R. Rocha Jr for valuable suggestions. RAM acknowledges support
from FAPESP (process number 98/16486-8). JV is grateful to CNPq (300707/93-2)
and FAPESP (01/01618-0) for partial financial support.}

\linespread{1}

\end{document}